\documentclass[preprint,journal]{vgtc}

\newcommand{\bstart}[1]{\vspace{0.8mm} \noindent{\textbf{#1:}}}

\newcommand{\bstartnvs}[1]{\noindent{\textbf{#1:}}}

\usepackage{xcolor}
\usepackage{xspace,xpunctuate}
\newcommand{\ie}{{i.e.,}\xspace}
\newcommand{\eg}{{e.g.,}\xspace}
\newcommand{\ea}{{et~al\xperiod}\xspace}

\newcommand{\etc}{{etc\xperiod}\xspace}

\definecolor{exOrg}{HTML}{f58233}
\definecolor{exBlue}{HTML}{327fc2}

\usepackage{amsmath,amssymb,amsthm}

\providecommand{\mat}[1]{\boldsymbol{\mathrm{\bf #1}}}%
\renewcommand{\vec}[1]{\boldsymbol{\mathrm{\bf #1}}}

\DeclareMathOperator{\hugeE}{\mbox{\huge\raise-0.3ex\hbox{E}}}
\DeclareMathOperator{\p}{\mathbb{P}}
\DeclareMathOperator{\hugep}{\mbox{\huge\raise-0.3ex\hbox{$\p$}}}

\newcommand{\RR}{\mathbb{R}}

\providecommand{\mA}{\ensuremath{\mat{A}}}
\providecommand{\mB}{\ensuremath{\mat{B}}}

\providecommand{\mD}{\ensuremath{\mat{D}}}

\providecommand{\mX}{\ensuremath{\mat{X}}}

\providecommand{\vx}{\ensuremath{\vec{x}}}

\ifpdf%                                % if we use pdflatex
  \pdfoutput=1\relax                   % create PDFs from pdfLaTeX
  \usepackage{graphicx}                % allow us to embed graphics files
  \DeclareGraphicsExtensions{.pdf,.png,.jpg,.jpeg} % for pdflatex we expect .pdf, .png, or .jpg files
\else%                                 % else we use pure latex
  \usepackage{graphicx}                % allow us to embed graphics files
  \DeclareGraphicsExtensions{.eps}     % for pure latex we expect eps files
\fi%

\graphicspath{{figures/}{pictures/}{images/}{./}} % where to search for the images
\usepackage[normalem]{ulem}
\usepackage{microtype}                 % use micro-typography (slightly more compact, better to read)
\PassOptionsToPackage{warn}{textcomp}  % to address font issues with \textrightarrow
\usepackage{textcomp}                  % use better special symbols
\usepackage{mathptmx}                  % use matching math font
\usepackage{times}                     % we use Times as the main font
\usepackage{cite}                      % needed to automatically sort the references
\usepackage{tabu}                      % only used for the table example
\usepackage{booktabs}                  % only used for the table example
\onlineid{0}

\vgtccategory{Research}
\vgtcpapertype{Theoretical \& Empirical}

\title{An Automated Approach to Reasoning About Task-Oriented Insights in Responsive Visualization}

\author{Hyeok Kim, Ryan Rossi, Abhraneel Sarma, Dominik Moritz, and Jessica Hullman}
\authorfooter{
\item
 Hyeok Kim, Abhraneel Sarma, and Jessica Hullman are with Northwestern University. E-mail: hyeokkim2024@u.northwestern.edu, abhraneelsarma2024 @u.northwestern.edu, jhullman@northwestern.edu.
\item
 Ryan Rossi is with Adobe Research. E-mail: ryrossi@adobe.com.
\item
 Dominik Moritz is with Carnegie Mellon University. E-mail: domoritz@cmu.edu.
}

\shortauthortitle{Kim \MakeLowercase{\textit{et al.}}: An Automated Approach to Reasoning About Task-Oriented Insights in Responsive Visualization}
\abstract{
Authors often transform a large screen visualization for smaller displays through rescaling, aggregation and other techniques when creating visualizations for both desktop and mobile devices (\ie~responsive visualization). However, transformations can alter relationships or patterns implied by the large screen view, requiring authors to reason carefully about what information to preserve while adjusting their design for the smaller display. We propose an automated approach to approximating the loss of support for task-oriented visualization insights (identification, comparison, and trend) in responsive transformation of a source visualization. We operationalize identification, comparison, and trend loss as objective functions calculated by comparing properties of the rendered source visualization to each realized target (small screen) visualization. To evaluate the utility of our approach, we train machine learning models on human ranked small screen alternative visualizations across a set of source visualizations. We find that our approach achieves an accuracy of 84\% (random forest model) in ranking visualizations. We demonstrate this approach in a prototype responsive visualization recommender that enumerates responsive transformations using Answer Set Programming and evaluates the preservation of task-oriented insights using our loss measures. We discuss implications of our approach for the development of automated and semi-automated responsive visualization recommendation.
} % end of abstract


\CCScatlist{ % not used in journal version
 \CCScat{K.6.1}{Management of Computing and Information Systems}%
{Project and People Management}{Life Cycle};
 \CCScat{K.7.m}{The Computing Profession}{Miscellaneous}{Ethics}
}

\vgtcinsertpkg

\begin{document}

%% The ``\maketitle'' command must be the first command after the
%% ``\begin{document}'' command. It prepares and prints the title block.

%% the only exception to this rule is the \firstsection command
\firstsection{Introduction}

\maketitle
\begin{figure}[t]
    \centering
    \includegraphics[width=\columnwidth]{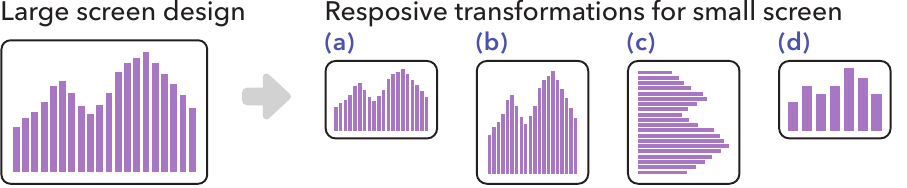}
    \caption{Example responsive transformations for small screen generated from a large screen design: (a) proportionate rescaling, (b) disproportionate rescaling, (c) transposing axes, and (d) increasing bin size.}
    \label{fig:responsive-example}
\end{figure}

Visualization authors often transform their designs to accommodate different audience, styles, or display types. 
For example, authors might simplify charts for audiences with different graphical literacy, altering information conveyed in the new design~\cite{cui2006}.
Authors of responsive visualizations create multiple designs for different screen sizes and interactivity~\cite{hoffswell2020,kim2021}. 
However, transforming visualizations may invoke trade-offs between desirable design criteria. For instance, in \autoref{fig:responsive-example}, proportionate rescaling (a) makes it harder to compare values along the \textit{y}-axis due to the reduced absolute height, while increasing the relative height (b) or transposing (c) can distort the shape of the represented distribution. 
Increasing the bin size (d) reduces possible comparisons and a viewer's ability to see distributional detail.

Unfortunately, design trade-offs make it difficult to reason about preserving takeaways or ``insights'' under visualization design transformations.
Designers may need to iteratively try out different combinations of strategies like those in \autoref{fig:responsive-example} and compare them with the original design~\cite{hoffswell2020}.
Kim~\ea~\cite{kim2021} identify trade-offs in designing responsive visualization where authors try to strike a balance between maintaining graphical density (\ie~an appropriate number of marks per unit screen size) and the preservation of a user's ability to arrive at certain insights.
For example, authors need to decide either to preserve distributional characteristics with higher visual density by rescaling proportionately (a) or to adjust visual density while losing distributional details by changing the bin size (d). Currently these decisions are made manually.

We contribute an automated approach to approximating the amount of change to task-oriented insights---insights that viewers are likely to be able to obtain from a visualization by performing visual tasks---under design transformation.
We define measures that approximate a visualization's support for three low-level visual analytic tasks discussed in the literature~\cite{amar2005,Brehmer2013}: the viewer's ability to identify a datum, to compare pairs of data points, and to perceive a multivariate trend.
We demonstrate the use of our measures for choosing between alternative transformations of a source large screen visualization in a responsive design context.
We provide a prototype automated design recommender for responsive visualization that enumerates responsive design transformations based on an input source view and reasons about changes in task-oriented insights from the source design to each transformed design.
Our recommender supports scatterplots, bar charts, line graphs, and heatmaps with position, color, size, and shape encoding channels.

We train and test different machine learning (ML) models based on our measures to evaluate their utility for automatically ranking small screen visualization design alternatives given a large screen view. 
We achieve up to 84\% accuracy (via a random forest model) in ranking a set of responsive transformations across a set of six source large screen views spanning different encoding channels. 
Models trained with our measures outperform two baseline models based on simple heuristics related to chart size changes (59\%) and transposing of axes (63\%).
We discuss implications of our work for future research, including recommender-driven responsive visualization authoring and generalizing our approach to further visualization design domains such as simplification and style transfer.

\section{Related Work}\label{sec:rw}
\subsection{Responsive Visualization and Design Transformations}\label{sec:rw_needs}
We use \textit{visualization design transformation} to refer to the transformation of a source visualization specification to a new visualization specification intended to better achieve certain context-specific constraints. These might be screen size limitations for responsive visualization, audience-related constraints in visualization simplification or audience retargeting (\eg~\cite{Bottinger2020,Johnson2014}), or style constraints in style transfer~\cite{Harper2014,harper2017}, for example. 
Visualization design transformation differs from creating multiple different views from the same dataset (\eg~a visualization sequence or dashboard) in that transformations of an original source view typically are intended to preserve many properties of the source while changing select properties.

Prior work on responsive visualization, which tends to focus on Web-based communicative visualization, or scalable visualization~\cite{cook2005illuminating} more broadly, emphasizes the importance of maintaining intended takeaways between source and transformed views.  
Analyzing 378 responsive visualization pairs on desktop and mobile devices, Kim~\ea~\cite{kim2021} identify density-message trade-offs in responsive visualization where authors need to balance adjusting visual density or complexity for different screen types while maintaining patterns, trends or other important information conveyed in the source view. 
Focusing on maintaining key information at different scales, earlier work on visualization resizing introduces algorithms that repeatedly remove the pixels determined to be least important~\cite{Giacomo2015} and iteratively minimize scaling in more salient regions~\cite{Wu2013}, for example.
We extend prior approaches by proposing approximation methods for task-oriented visualization insights.

Defaulting to simpler views over complex, over-encoded plots is often recommended when exploring or publicizing complex data~\cite{kelleher2011}. 
Authors accomplish this through data-level transformations, such as data abstraction, clutter reduction, filtering, or clustering. 
For example, data abstraction studies have attempted to enhance the simplicity of a view while preserving original structure or insights (\eg~aggregating a large movement dataset~\cite{Adrienko2011}, using interactive dimensionality reduction~\cite{Johansson2009}, using hierarchical aggregation~\cite{Elmqvist2010}, and measuring the quality of an abstraction~\cite{cui2006}).
\subsection{Visualization Recommendation}\label{sec:rw_insight_driven}
We discuss two approaches in visualization recommendation---insight-based and similarity-based---that are relevant to our goal of approximating changes in task-oriented insights.
Prior work on visualization recommendation employs statistical calculations to characterize properties of a visualization thought to relate to the insights a user can draw from it.  
Often these `insights' are intended to capture how well a user can perform analytic tasks, such as recognizing trends or identifying and comparing data points.
Tang \ea~\cite{tang2017topk} suggest detecting `top-k insights' from data using statistical significance testing (\eg~low $p$-value of a linear regression coefficient for slope insight).
Similarly, Foresight~\cite{demiralp2017}, Data\-Site~\cite{cui2019}, and Voder~\cite{Srinivasan2019} use statistics calculated on the data, such as correlation coefficient and interquartile range, and recommend visualization types predicted to better support extracting such information.
However, statistics on data are invariant for views sharing the same data set and hence of limited use for comparing different ways of visualizing the same underlying data.
Our work instead considers statistics calculated on the rendered visualization. 

Several prior visualization recommenders model similarity between views, but assume a scenario where the underlying dataset changes.
GraphScape~\cite{kim2017graphscape} offers a view similarity model that assigns costs to visualization pairs that are intended to approximate the cognitive cost of transitioning from one view to another in a visualization sequence. 
GraphScape applies an a priori cost model in which data transformation (\eg~binning, modifying scales) is always less costly than changes in encoding. Hence, filtering data has a lower cost than transposing axes. 
However, filtering operations like removing a bar from a bar chart or rescaling a \textit{y}-axis can significantly change the presumed ``take-aways'' of a chart (\eg~\cite{gelman2020,hofman2020visualizing}). 
The space of transformations covered by GraphScape also does not include view size transformations, so it cannot assign costs to changes in aspect ratio common to responsive visualization.
Although Dziban~\cite{dziban} extends GraphScape to suggest a view that is `anchored' to the previous view for an exploratory data analysis process, it also assumes different subsets of data between the previous and current views and focuses more on similar chart encodings than on preserving task-oriented insights.

\subsection{Comparing Visual Structure by Processing Signal}\label{sec:rw_signal_processing}
Signal processing-based approaches analyze the underlying visual or perceptual structure of a visualization to enable multi-scale visualizations (\ie~providing different insights at different scales) and to enhance visualization effectiveness.
Prior work has attempted to enable multi-scale views through perceptual organization analysis of a information graphic at each scale~\cite{Wattenberg2003,Wattenberg2004} and hybrid-image visualization that displays different aggregation levels at different viewing distances~\cite{Isenberg2013}, for example.
Signal processing approaches have also been applied to improve the effectiveness of a visualization, for instance, by measuring the difference between the visual salience of a representation and salience of signals in data~\cite{janicke2010,Kindlmann2014}, comparing kernel density estimations between a LOESS curve and different representations~\cite{Wang2018}, and extending a structural similarity index for image compression to data visualization~\cite{Veras2020}.
Signal processing-based approaches have typically been applied to single views, and are generally confined to a predefined set of marks and visual variables (\eg~a line chart, a scatterplot), restricting their applicability for settings like ours. 

\section{Problem Formulation}\label{sec:scenario}
\begin{figure}[t]
    \centering
    \includegraphics[width=\columnwidth]{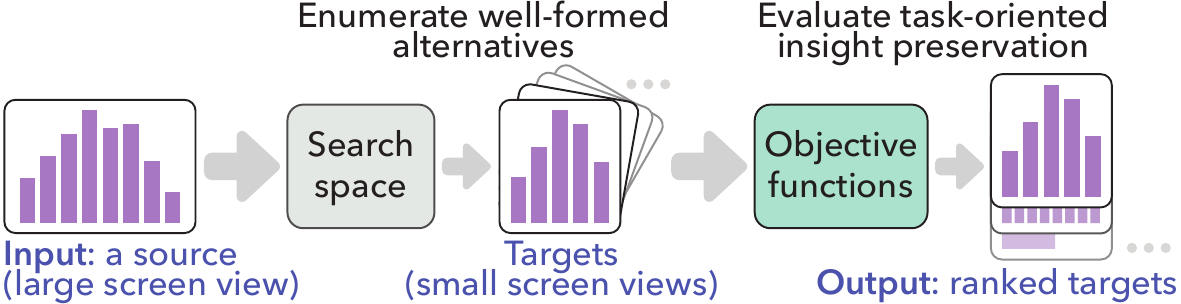}
    \caption{A pipeline for a responsive visualization recommender}
    \label{fig:pipeline}
\end{figure}

We propose formulating responsive visualization as a search problem from an input source view to transformed target views, following the characterization proposed by Kim~\ea~\cite{kim2021}. 
Consider a recommender that takes a source desktop view as input and returns a ranked set of targets as illustrated in \autoref{fig:pipeline}.
The first step in creating such a recommender is to define a search space that can enumerate well-formed responsive targets. 
To generate useful target views from a source (large screen) visualization, a search space should cover common transformation strategies in responsive visualization, such as rescaling, aggregating, binning, and transposing~\cite{kim2021,hoffswell2020}.

After enumerating target views, a responsive visualization recommender should evaluate how well each target preserves certain information or ``insights.'' 
While the term insight can be overloaded~\cite{zgraggen2018investigating}, a relatively robust way to define insights comes from typologies for describing visualization judgments or patterns~\cite{amar2005, Brehmer2013}. These typologies suggest defining insights around common low-level visual analysis tasks like identifying and comparing data. In an automated design recommendation scenario, these task-oriented insights can be approximated by objective functions (\ie~loss measures) that capture support for common tasks, applied  to both the source and target view.
Finally, the recommender returns the set of target designs based on how well they minimize these loss measures. 
We formalize this problem and motivate and define three loss measures that we call \emph{task-oriented insight preservation measures}. In \autoref{sec:prototype}, we describe a prototype visualization recommender in which we implemented the approach. 

\subsection{Notation}\label{sec:notations}
\begin{figure}[t]
    \centering
    \includegraphics[width=\columnwidth]{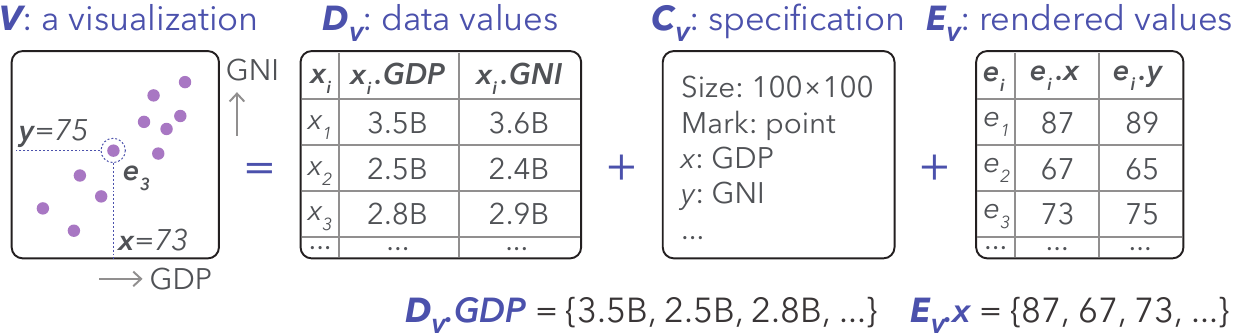}
    \caption{Our notation for a visualization. Rendered values are defined in the space implied by the visual variable (\eg~pixel space for position or size, color space for color).}
    \label{fig:notation-vis}
\end{figure}

We define a visualization (or a view), $V$, as a three tuple
\begin{equation}
    V = [D_V, C_V, E_V],
    \label{eq:view}
\end{equation}
where $D_V$ is the data used in $V$, $C_V$ is a visualization specification (defining encodings, chart size, mark type, \etc), and $E_V$ is a set of rendered values that we compute our measures on.
For example, suppose a bivariate data set with \textit{GDP} and \textit{GNI} fields (\ie~$D_V=\{x_1, x_2, \dots, x_n\}$, where $x_i=(x_i.\textit{GDP}, x_i.\textit{GNI})$).
$C_V$ maps GDP and GNI to \textit{x} and \textit{y} positions of point marks, respectively, producing a scatterplot.
The corresponding set of rendered values is a set of Cartesian coordinates on the XY-plane (\ie~$E_V=\{e_1, e_2, \dots, e_n\}$, where $e_i=(e_i.x, e_i.y)$ is the tuple of rendered values for $x_i$). 
Similarly, for a data set containing a field $CO2$ (emission) that is mapped to $color$, $e_i.\textit{color}$ would correspond to the rendered value of $x_i.\textit{CO2}$.
For brevity, we define $D_V.\textit{field}$ as a vector of \textit{field} values and $E_V.\textit{channel}$ as a vector of rendered values in \textit{channel}.
Our notation is also illustrated in \autoref{fig:notation-vis}

Given a source view $\mathbb{S}$ and a transformation (or target) $\mathbb{T}$, we represent the loss of insight type $M$ from $\mathbb{S}$ to $\mathbb{T}$ as below:
\begin{equation}
    \textit{Loss}(\mathbb{S} \rightarrow \mathbb{T}; M)
    \label{eq:loss}
\end{equation}
For example, $\textit{Loss}(\mathbb{S} \rightarrow \mathbb{T}; \text{Trend})$ indicates trend loss from $\mathbb{S}$ to $\mathbb{T}$.

\section{Task-oriented Insight Preservation Measures}\label{sec:loss_measures}
High level criteria for preserving task-oriented insights of a visualization include preserving datum-level information, maintaining comparability of data points, and preserving the aggregate features~\cite{Brehmer2013}.
We use these distinct classes of information to define task-oriented insight loss measures for approximating how well a responsive transformation preserves support for low-level tasks of identifying data, comparing data, and identifying trend.
Our goal is to define a small set of measures that capture important types of low-level tasks a designer might wish to preserve in responsive transformation.
Each measure should be distinct (\ie~mostly independent of the others) and should improve accuracy when combined with the others (such as through regression or ML modeling) to predict human judgments about how visualization transformations rank.
Together, the measures should outperform reasonable baseline approaches based on simple heuristics. 
While chosen to cover three important classes of low-level analytic task, the measures we describe are not meant to be exhaustive, as there are many ways one could approximate support for task-oriented insights.

\subsection{Identification Loss}\label{sec:loss_identification}
\begin{figure}[b]
    \centering
    \includegraphics[width=\columnwidth]{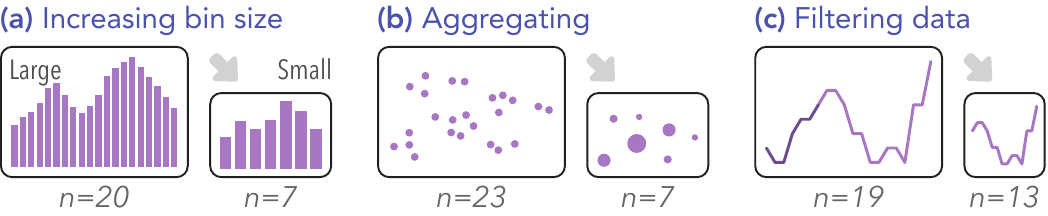}
    \caption{Responsive transformations that may cause identification loss.}
    \label{fig:motivation-identify}
\end{figure}

Responsive visualization strategies often alter the number of visual attributes of marks that viewers can identify (affecting a low-level identification task~\cite{amar2005,Brehmer2013}).
As illustrated in \autoref{fig:motivation-identify}, when the number of bin buckets of a histogram is decreased in a mobile view (a), each bar encodes more information on average than in the desktop view, such that some information about the distribution is lost.
Similarly, strategies to adjust graphical density, like aggregating distributions (b) and filtering certain data (c), also reduce the number of identifiable attributes. 
We use \textit{identification loss} to refer to changes to the identifiability of rendered values between a source view and a target.

Information theory, and in particular Shannon \textit{Entropy} (entropy, hereafter) captures the information in a signal by measuring the minimum number of bits needed to encode it~\cite{shannon1948}. 
Given a random variable $X$, entropy is defined as $H(X) = - \sum_{x \in X}{P(x) \log_2{P(x)}}$. 
Applying this to visualization, suppose that for a source visualization $\mathbb{S}$, a vector for data field $f$, $D_\mathbb{S}.\textit{f} = \{x_1.\textit{f},\dots,x_n.\textit{f}\}$, is mapped to an encoding channel $c$. 
The corresponding rendered values $E_\mathbb{S}.c=\{e_1.c,\dots,e_n.c\}$ compose a random variable $U_\mathbb{S}.c$ that takes the set of unique values of $E_\mathbb{S}.c$ as its outcome space, where the probability of $U_\mathbb{S}.c$ taking $x$ is defined as the relative frequency of $x$ in $E_\mathbb{S}.c$, formalized as
\begin{equation}
    P(U_\mathbb{S}.c = x)=\textit{Count}_{i}(e_i.c=x) / n 
    \label{eq:entropy_probability}
\end{equation}
\begin{equation}
    H(E_\mathbb{S}.c)= -\sum_{x}{P(U_\mathbb{S}.c = x) \log_2{P(U_\mathbb{S}.c = x)}}
    \label{eq:entropy_entropy}
\end{equation}

We can similarly compute the probabilities of rendered values, $P(U_\mathbb{T}.c)$, and the entropy of an encoding channel, $H(E_\mathbb{T}.c)$, for a target view $\mathbb{T}$. 
Finally, we can calculate the identification loss for the channel as the absolute difference in entropy (\ie~$|H(E_{\mathbb{S}}.c) - H(E_{\mathbb{T}}.c)|$), where 0 difference is the identity.
The final identification loss from $\mathbb{S}$ to $\mathbb{T}$ is the sum of absolute differences in entropy for each encoding channel $c$ between the two views:
\begin{equation}
    \textit{Loss}(\mathbb{S} \rightarrow \mathbb{T}; \text{Identification}) = 
    \sum_{c}|H(E_{\mathbb{S}}.c) - H(E_{\mathbb{T}}.c)|,
    \label{eq:loss_entropy}
\end{equation}
\subsection{Comparison Loss}\label{sec:loss_comparison}
\begin{figure}[b]
    \centering
    \includegraphics[width=\columnwidth]{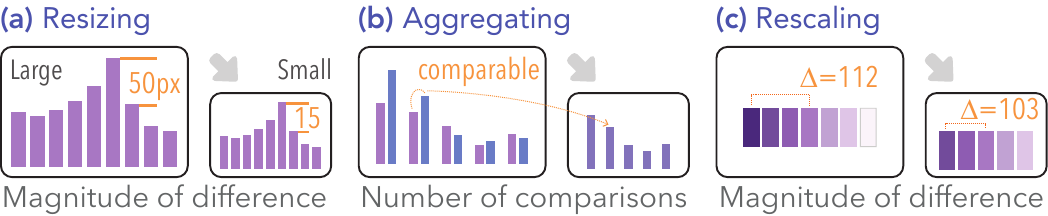}
    \caption{Responsive transformations that may cause comparison loss.}
    \label{fig:motivation-compare}
\end{figure}

Responsive transformations like resizing or scaling a view or aggregating data can alter the number of possible data comparisons that a user can make and how perceptually difficult they are (affecting a low-level comparison task~\cite{amar2005,Brehmer2013}). 
For instance, in \autoref{fig:motivation-compare}, resizing (a) diminishes the magnitude of difference between two highlighted data points in the small screen design.
In a mobile design with aggregation (b), viewers are no longer able to make each comparison that is available in the large screen view.
This motivates estimating how similarly viewers are able to discriminate between pairs of points in a target view compared to the source view, which we refer to as \textit{comparison loss}.

Empirical visualization studies (\eg~\cite{szafir2018,kim2018}) often operationalizes accuracy as the viewer's ability to perceive relationships between pairs of values.
While simpler scalar statistics like a sum or mean might suffice under some transformations, a method that preserves the distribution of distances will be more robust to transformations that change the number of data points or scales (\eg~log-scale).
We operationalize comparison loss as the difference in pairwise discriminability, measured using Earth Mover's Distance (EMD), between the source and a target in each encoding channel used in a visualization:
\begin{equation}
    \textit{Loss}(\mathbb{S} \rightarrow \mathbb{T}; \text{Comparison}) = 
    \sum_{c}EMD(B_{\mathbb{S}}.c, B_{\mathbb{T}}.c'),
    \label{eq:loss_emd}
\end{equation}
where $B_{\mathbb{S}}.c$ and  $B_{\mathbb{S}}.c'$ are the \textit{discriminability distributions} of the source and target views in encoding channel $c$ and $c'$, respectively, that encode the same data field.

Given a source visualization $\mathbb{S}$, we define the discriminability distribution $B_{\mathbb{S}}.c$, of an encoding channel $c$ for a view $\mathbb{S}$, as the set of distances between each pair of rendered values ($E_\mathbb{S}.c$) of $\mathbb{S}$ in terms of $c$.
This is formalized as
\begin{equation}
    B_{\mathbb{S}}.c = \{d_c(e_i.c, e_j.c) : e_i.c, e_j.c \in E_\mathbb{S}.c\},
    \label{eq:discriminability_distribution}
\end{equation}
where $d_c(\cdot,\cdot)$ is a distance metric for the encoding channel $c$.

\bstart{Distance metrics} 
Ideally, comparison loss should account for differences in how well visual channels support perception of numerical values. 
Informed by visual perception models, we select several distance metrics intended to provide a rough proxy of the perceptual difference between two visual signals.
While visual variables can have interaction effects~\cite{brychtova2017,smart2019,szafir2018}, for simplicity in demonstrating our approach, we limit our use of perceptual distance metrics to encoding channel specific measures. However, as the state-of-the-art in predicting effects of visual variable interactions develops, our approach could be amended to consider combinations. 

For position channels, we use the absolute difference between two position values (in pixel space), as human vision is highly accurate in discriminating positions according to Stevens' power law~\cite{stevens1957psychophysical,stevens2017psychophysics} and empirical studies~\cite{heer2009,heer2010}:
\begin{equation}
    d_{\textit{position}}(e_i.\textit{position}, e_j.\textit{position}) = |e_i.\textit{position} - e_j.\textit{position}|
    \label{eq:distance_position}
\end{equation}
We measure distance in a size channel using the absolute difference between two size values (in pixel) raised to the estimated Stevens' exponent of 0.7~\cite{stevens1957psychophysical,stevens2017psychophysics}: 
\begin{equation}
    d_{\textit{size}}(e_i.\textit{size}, e_j.\textit{size}) = |e_i.\textit{size} - e_j.\textit{size}|^{0.7}
    \label{eq:distance_size}
\end{equation}

We calculate the Euclidean distance in the perceptual color space CIELAB~\cite{fairchild2004color} (CIELAB 2002):
\begin{multline}
    d_{\textit{color}}(e_i.\textit{color}, e_j.\textit{color}) = 
    \\ \sqrt{(e_i.L-e_j.L)^2 + (e_i.a-e_j.a)^2 + (e_i.b-e_j.b)^2}
    \label{eq:distance_color},
\end{multline}
where $L$, $a$, and $b$ represent $L*$, $a*$, and $b*$ in CIELAB space.

Lastly, for shape encodings, we employ a perceptual kernel~\cite{demiralp2014}, a (symmetric) matrix of pairwise distances between visual attributes.
The $i,j$-th element in the perceptual kernel for shape is the empirical probability of discriminating shape $i$ from shape $j$ based on an online crowdsourced experiment in which workers completed a triplet discrimination task where they chose the most dissimilar shape out of three shapes. 
Formally, our shape distance metric can be stated as:
\begin{multline}
    d_{\textit{shape}}(e_i.\textit{shape}, e_j.\textit{shape}) = \\ P(e_i.\textit{shape} \text{ is discriminated from } e_j.\textit{shape})
    \label{eq:distance_shape}
\end{multline}

\bstartnvs{Comparing discriminability distributions}
To quantify the discrepancy between the discriminability distributions of encoding channel $c$ and $c'$ (mapping the same field) for the source $\mathbb{S}$ and target $\mathbb{T}$ (\ie~$B_{\mathbb{S}}.c$ and $B_{\mathbb{T}}.c'$, respectively), we compute Earth Mover's Distance~\cite{Villani2009} (EMD or Wasserstein distance).
We use EMD, which measures the minimum cost to transform a distribution to another distribution, because it is non-parametric, symmetric, and unbounded.
An EMD of 0 is the identity, and the greater the EMD is, the more different the two distributions are.
Thus, the comparison loss between the source view $\mathbb{S}$ and a target view $\mathbb{T}$ is the sum of the EMD between their discriminability distributions in each encoding channel, formalized in \autoref{eq:loss_emd}.
\begin{figure}[b]
    \centering
    \includegraphics[width=\columnwidth]{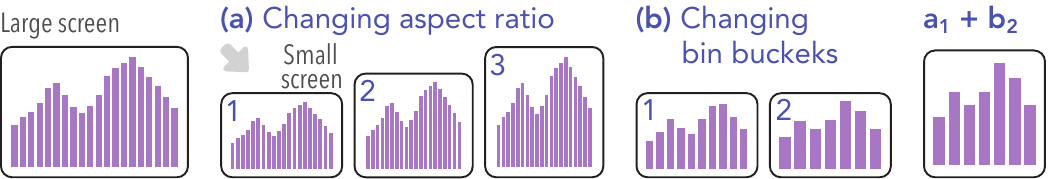}
    \caption{Responsive transformations motivating trend loss.}
    \label{fig:motivation-trend}
\end{figure}

\subsection{Trend Loss}
Responsive transformations like disproportionate rescaling and changes to binning may impact the implied relationship (or trend) between two or more variables represented in a target view compared to the source view (affecting low-level trend identification~\cite{Brehmer2013}).
As shown in \autoref{fig:motivation-trend}, different aspect ratios can alter the magnitude of the slope of a trend, and modifying bin size affect the amount of distributional information available.
We use \textit{trend loss} to refer to changes in the implied trend from the source to a target.

\begin{figure}[!b]
    \centering
    \includegraphics[width=\columnwidth]{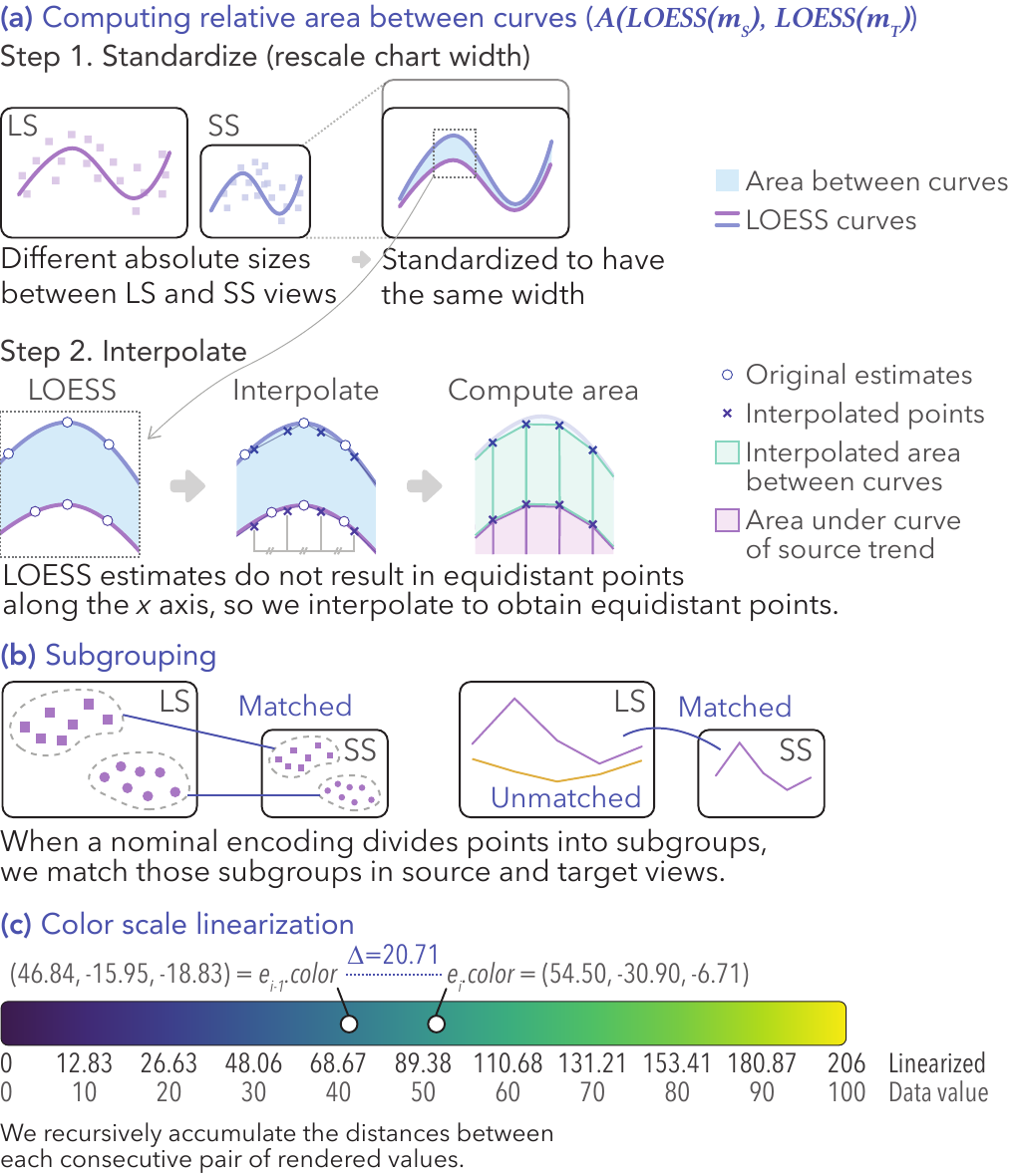}
    \caption{Components of computing trend loss. (a) Calculating area between curves by standardizing chart size and interpolating break points. (b) Dividing and matching subgroups. (c) Linearizing color scale. LS is large screen, and SS is small screen. 
    }
    \label{fig:trend-figure}
\end{figure}

\begin{figure*}[t]
    \centering
    \includegraphics[width=\textwidth]{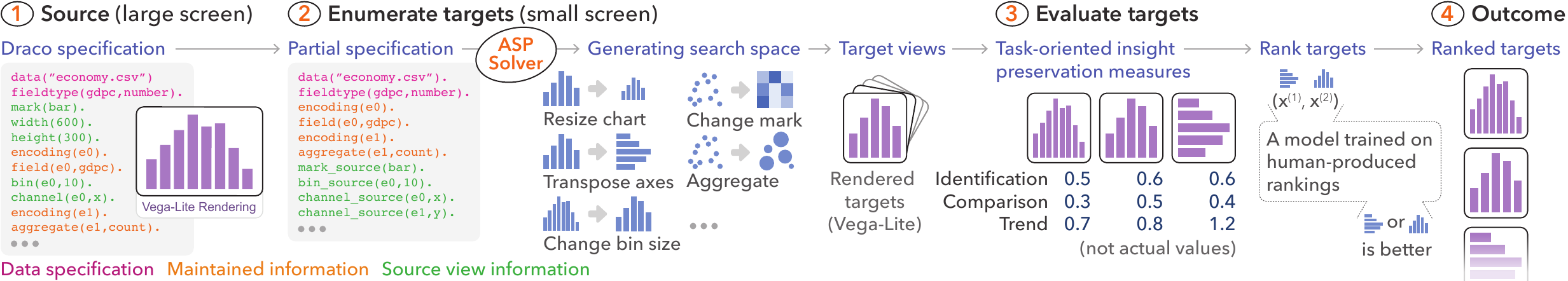}
    \caption{Prototype pipeline. (1) The full specification of an input source view in ASP. (2) Enumerating targets by extracting a partial specification of the source view and generating a search space using an ASP solver. (3) Evaluating targets by computing our loss measures and ranking them using a model trained on human-produced rankings. (4) Ranked targets. }
    \label{fig:prototype}
\end{figure*}

To capture representative data patterns while avoiding influences of noise, our trend loss first estimates trend models between the source and target views using LOESS.
We then compare the area (or volume) of the estimated trends because it is more sensitive to details that simpler methods (\eg~the difference between regression coefficients) might ignore.
We define trend models for the quantitative encoding channels in our scope (position, color and size): 
\begin{itemize}
    \item $e_y \sim e_x$: a 2D trend of \textit{y} on \textit{x} as appears in a simple scatterplot, line chart, or bar graph.
    \item $e_{color} \sim e_x + e_y$: a 3D trend of color on \textit{x} and \textit{y} like a heatmap or a scatterplot with a continuous color channel
    \item $e_{size} \sim e_x + e_y$: a 3D trend of size on \textit{x} and \textit{y} (\eg~a scatterplot with a continuous size encoding)
\end{itemize}

After calculating trend models for a source and target, we can define trend loss as the sum of the relative area between curves (or volume between surfaces) of the estimated trends in each trend model ($m$).
This is formalized as:
\begin{equation}
    \textit{Loss}(\mathbb{S} \rightarrow \mathbb{T}; \text{Trend}) = 
    \sum_{m}A(LOESS({m_\mathbb{S}}), LOESS({m_\mathbb{T}}))
    \label{eq:loss_trend}
\end{equation}
where $A$ stands for the relative area between curves (ABC) between the source and target trends ($m_\mathbb{S}$ and $m_\mathbb{T}$), normalized by dividing by the area under the curve of the source trend for a 2D model. 
For a 3D model, $A$ is the relative volume between surfaces (VBS), which is the VBS of the source and target trends divided by the volume under the surface of the source trend.

We estimate the trend models using LOESS regression~\cite{cleveland1979} as it is non-parametric.
We use uniform weights and bandwidth of 0.5~\cite{cleveland1979}.
LOESS regression returns an estimate at each observed value of the independent variable(s) (as an array of coordinates): an estimated curve for a 2D model and an estimated surface for a 3D model. 
Thus, when source and target views have different chart sizes or different sets of rendered values for the independent variable(s), it is difficult to directly compare the LOESS estimations.
As shown in \autoref{fig:trend-figure}a, we first standardize the chart sizes of two views by rescaling an estimated LOESS curve or surface in a target view to have the same chart width with the source. 
Then, we interpolate the LOESS curve to have equal distances between two consecutive coordinates for a 2D model (\autoref{fig:trend-figure}b).
We interpolate on 300 breakpoints in a 2D model by default, where one breakpoint corresponds to one to three pixels in many Web-based visualizations.
For a 3D model, we interpolate 300 $\times$ 300 breakpoints from a LOESS surface in a similar way. 
Given these interpolations for the LOESS curves (or surfaces) in the source and target, we obtain the ABC (or VBS) segment at each breakpoint.

\bstart{Subgroups}
When a nominal variable encoded by color or shape divides the data set into subgroups, viewers might naturally consider each subgroup's trend independently. 
To distinguish trends implied by subgroups, we first identify and match subgroups which occur in both the source and target views by looking at their nominal data values, as depicted in \autoref{fig:trend-figure}b.
Then, we compute the relative ABC (or VBS) of each subgroup and combine them by taking their average.

\bstart{Color scale linearization}
Although a continuous color scale encodes a unidimensional vector, color is often modeled on a multi-dimensional space (\eg~RGB, CIELAB), which makes it complex to estimate a LOESS surface. 
Similar to how common color schemes such as \textit{viridis} or \textit{magma} are designed to be perceptually uniform by keeping equi-distance in a perceptual color space between two consecutive color points~\cite{mpl}, we can make use of the Euclidean distance between rendered color values in CIELAB to linearize a 3D color scheme.
Specifically, we recursively accumulate the distances between each consecutive pair of rendered values to create a unidimensional vector.
In \autoref{fig:trend-figure}c, we show how the linear value of $i$-th color point is computed from that of $i-1$-th point; we take the calculated value of the $i-1$-th point and add to it the distance between the $i-1$-th and $i$-th points. 
The first color point is assigned as zero.

\begin{figure*}[ht]
    \centering
    \includegraphics[width=\textwidth]{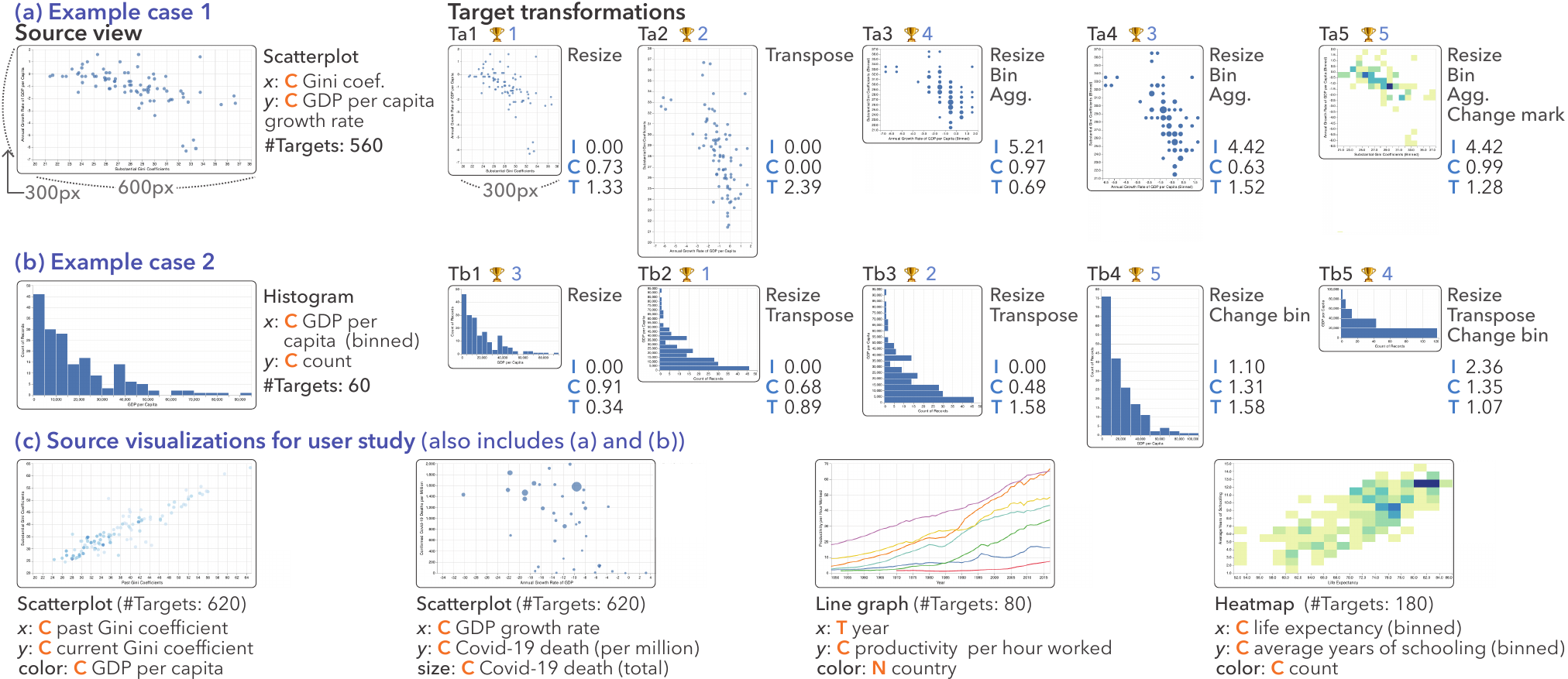}
    \caption{(a, b) Example target transformations enumerated by our prototype responsive visualization recommender (total size of search space per source given as \#Targets).
    \includegraphics[scale=1.2]{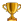} indicates rankings of each five targets per source view predicted by our best model (see \autoref{sec:results}).
    (c) Source visualizations for our user study (also includes a and b). Sources views have width of 600px and height of 300px. The width of targets is fixed as 300px. Data sets are from \textit{Our World in Data}~\cite{owid-economy,owid-health,owid-life}.
    \textbf{\textcolor{exOrg}C}ontinuous, \textbf{\textcolor{exOrg}N}ominal, \textbf{\textcolor{exOrg}T}emporal, \textbf{\textcolor{exBlue}I}dentification loss, \textbf{\textcolor{exBlue}C}omparison loss, and \textbf{\textcolor{exBlue}T}rend loss.}
    \label{fig:examples}
\end{figure*}

\section{Prototype Responsive Visualization Recommender}\label{sec:prototype}
To implement our task-oriented insight preservation measures, we developed a prototype responsive visualization recommender that enumerates and evaluates responsive designs (or targets).
As shown in \autoref{fig:prototype}, given an input source (large screen) view, our recommender first converts it to a partial specification, and then generates a search space of small screen targets based on the partial specification. 
We adopt the desktop-first approach that visualization authors have described using~\cite{hoffswell2020,kim2021}.
Finally, the recommender computes our measures between the source view and each target to rank those targets using an ML model trained on human-labeled rankings. 

\subsection{Enumerating Target Views}
To enumerate target views, we need a formal grammar for representing visualization specifications and formulating a search space.
We use Answer Set Programming (ASP)~\cite{Brewka2011}, particularly by modifying Draco~\cite{draco}.
ASP is a declarative programming language for complex search problems (\eg~satisfiability problems) that encodes knowledge as facts, rules, and constraints. Rules generate further facts, and constraints prevent certain combinations of facts.
Formalized in ASP, for example, Draco has a rule that if an encoding is binned, then it is discrete, and a constraint that disallows logarithmic scale on a discrete encoding~\cite{draco}.
A constraint solver then solves an ASP program (the partial specification of a source view and our search space), returning stable sets of non-conflicting facts (enumerated target views with different transformation strategies).
We use Clingo~\cite{gebser2014,gebser2011} as our solver.

\bstart{Converting to a partial specification} Our recommender converts the full specification of an input source view to a partial specification to allow applying responsive transformation strategies.
We maintain the data specification (data file, data field definitions, and the number of rows) and encoding information (\eg~count aggregation, association of data field) that are not changed under transformation.
We indicate the rest of the specification (mark type, chart size, and encoding channels) as information about the source view to constrain responsive transformation strategies (\eg~constraining possible mark type replacement, allowing for swapping position encodings for axis-transpose). 

\bstart{Generating a search space} Our goal in generating a search space is to produce a set of reasonable targets that a responsive visualization author might consider given a source view.
We generate a search space by automatically applying responsive visualization transformations recently observed in an empirical study of common responsive visualization design strategies~\cite{kim2021} to a source visualization. 
Our prototype implements rescaling, aggregation, binning, transposing, and select changes to marks and encodings. 
For rescaling, we fix the width of target views and vary heights, in the range from the height resulting from proportionate rescaling to the height that forms the inverse aspect ratio with an increment of 50\,px.
For example, if the source view has a width of 600\,px and a height of 300\,px (an aspect ratio of 2:1) and the width of target views is fixed at 300 px, then the height varies from 150\,px (2:1) to 600 (1:2) by 50\,px (\ie~150, 200, \dots, 550, 600\,px). 
Given a disaggregated source view, we generate alternatives by applying binning (max bin buckets of 25, 15, and 5) and aggregation (count, mean, median, sum) as graphical density adjustment strategies. 
We also generate alternatives by transposing axes (\ie~swapping \textit{x} and \textit{y} position channels).
Finally, in line with the observation of prior work that responsive visualization authors occasionally substituted mark types when adding an encoding channel for aggregation, we allow a mark type change in scatterplots from a point mark to a rectangle (heatmap).
We formulate these strategies in ASP format and add them to Draco~\cite{draco}.

\subsection{Evaluating and Ranking Targets}
To evaluate enumerated targets, we calculate our loss measures on rendered values after rendering source and target views using Vega-Lite~\cite{Satyanarayan2017vegalite}.
Then, we obtain rendered values, $E_V$, of a visualization $V$ by gleaning Vega~\cite{vega} states (a set of raw rendered values~\cite{vegaState}).
We implemented the loss measures in Python using SciPy~\cite{2020SciPy-NMeth}'s \texttt{stats.entropy} and \texttt{stats.wasserstein\_distance} methods for entropy and EMD, respectively.
To compute LOESS regression, we use the \verb|LOESS| package~\cite{pyLoess}.
Finally, to rank the enumerated targets, we combine the computed loss values by training ML models, which we detail in \autoref{sec:evaluation}.
We use ML models for ranking instead of formalizing them in ASP because our measures are not declarative (not rule-based).
\subsection{Examples}\label{sec:examples}
We introduce two example cases of transformations generated by our prototype and describe how our measures distinguish target views.

\subsubsection{Case 1: Simple scatterplot}
In the source scatterplot (\autoref{fig:examples}a), each point mark represents a country, and \textit{x} and \textit{y} positions encode Gini coefficients and annual growth rate of GDP per capita of different countries, respectively. 
The first example transformation (Ta1) is simple resizing. 
The second target view (Ta2) is transposed from the source view while keeping the size.
The third and fourth target views (Ta3 and Ta4) are resized, binned in \textit{x} and \textit{y} scales, and aggregated by count, so the size of each dot represents the number of data points in the corresponding bin bucket.
In the fifth target (Ta5), the mark type is changed from point to rectangle in addition to resizing, binning, and aggregating, and the color of each rectangle encodes the number of data points in that cell. 

Because Ta1 and Ta2 perfectly preserve the number of identifiable rendered values, identification loss is zero. 
Ta4 and Ta5 have more identifiable points than Ta3 (due to their smaller bin size), so they have smaller identification loss. 
While Ta1 has disaggregated values, Ta4 better preserves the distances between points in terms of position encoding, so it has smaller comparison loss.
Compared to the source view, the implied trend given \textit{x} and \textit{y} positions in Ta1 has a more similar slope and hence smaller trend loss than Ta2, whereas Ta2 preserves the differences in the position encodings, resulting in zero comparison loss.
Similarly, Ta3 has a smaller trend loss than Ta4 because Ta3 better preserves the visual shape of the distribution in the source view.

\subsubsection{Case 2: Histogram}
The source histogram in \autoref{fig:examples}b shows the distribution of GDP per capita of different countries. There are 23 bins along the \textit{x} axis and each bar height (\textit{y} position) represents the number of countries in the corresponding bin.
The first target view (Tb1) is resized.
The second and third target views (Tb2 and Tb3) are transposed with different resizing.
In the fourth and fifth target views (Tb4 and Tb5) bin sizes are changed from (23 to 10 and 5, respectively), with Tb5 transposed.

As Tb1, Tb2, and Tb3 have no changes in binning, they have zero identification loss, whereas Tb4 and Tb5 has greater identification loss proportional to their bin sizes.
While Tb1, Tb2, and Tb3 have the same binning, Tb3 has the most similar differences between bar heights and bar intervals in pixel space, so it has the smallest comparison loss among them.
Transposing axes (Tb3) better preserves the resolution for comparison (\ie~chart height and width), often resulting in the smaller comparison loss than other similarly transformed targets.
Tb5 has smaller trend loss than Tb4 as it shows a similar aspect ratio to the source view, though inverted, as implied by \textit{x} and \textit{y} positions.

\section{Model Training and Evaluation}\label{sec:evaluation}
A responsive visualization recommender should combine loss measures to rank a set of targets by how well they preserve task-oriented insights.
For our prototype recommender, we train machine learning models to efficiently combine our loss measures and rank enumerated targets. 
We describe training data collection, model specification, and results. 
\subsection{Labeling}
\begin{figure}[t]
    \centering
    \includegraphics[width=\columnwidth]{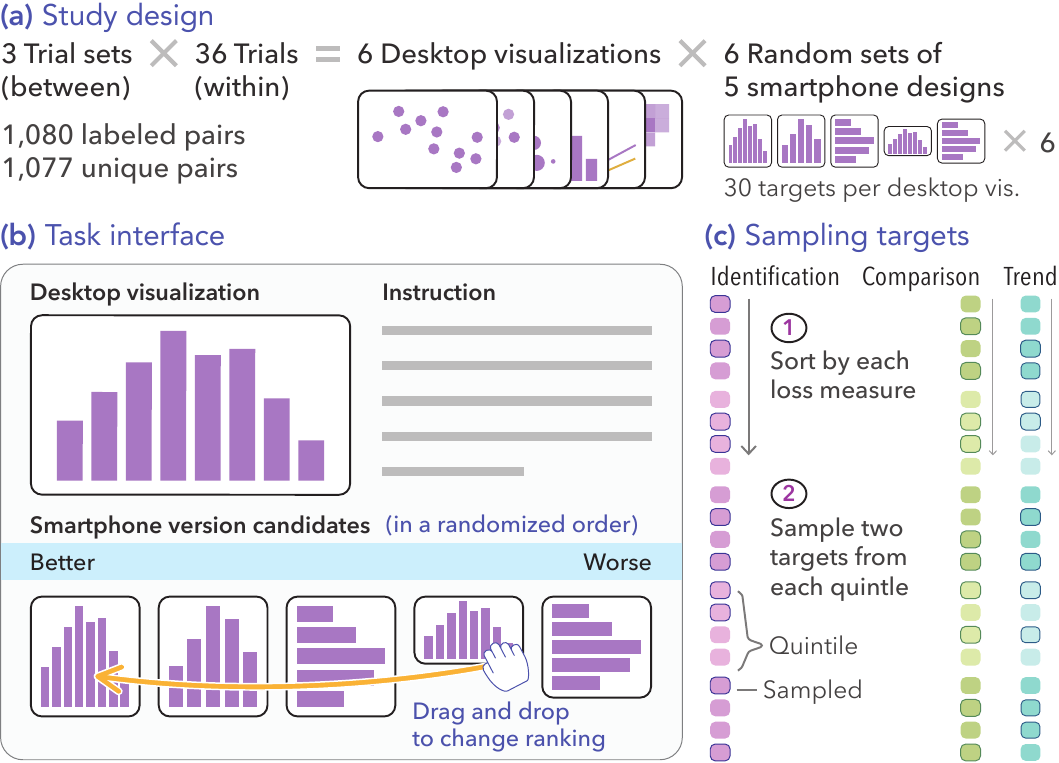}
    \caption{(a) Study design. (b) Task interface. (c) Quintile sampling of targets for task materials.}
    \label{fig:study-design}
\end{figure}

We obtained training and test data consisting of ranked target views for a set of source views using a Web-based task completed by nine visualization experts.
As shown in \autoref{fig:study-design}a, each labeler was assigned one out of three trial sets and performed 36 trials, with each trial asking them to rank five target transformations (small screen) given a source visualization (large screen).

\bstart{Task materials}
To create instances for labeling, we selected six desktop visualizations (source views) as shown in \autoref{fig:examples}c.
Our goal was to include different chart types, multiple encoding channels for identification and comparison losses, and different types of examples for trend loss (\eg~2D/3D models, subgroups, color scale linearization).
Our prototype generates 60 to 620 target transformations (2,120 in total) for these six source views.
We generated three sets of 30 target views per source view for labeling, using quintile sampling per preservation (loss) measure, to ensure relatively diverse sets of targets.
After sorting targets in terms of each of our three measures, we sampled two targets from each quintile of the top 100 targets per measure, as depicted in \autoref{fig:study-design}c.
We took the top 100 targets after inspecting the best ranked views per measure for each source view, to avoid labeling examples that might be obviously inferior.
Because identification loss is measured using entropy and is primarily affected by how data are binned, certain source views had fewer than five unique discrete values within top 100 targets. 
In this case, we proportionately sampled each discrete value.

After sampling 30 targets for a source view in a trial set, we randomly divided them into six trials (but fixed these trials between labeler in the same trial set), so we had 1,080 pairs (1,077 unique pairs\footnote{Each source view had 180 pairs, but the histogram source view with 60 transformations has 177 unique pairs.}) labeled by three people each.
We randomly assigned each trial set to labeler (3 trial sets (between) $\times$ 6 source views $\times$ 30 targets (within), \autoref{fig:study-design}a).

\bstart{Labelers}
All five authors, who have considerable background in visualization design and evaluation, and an additional convenience sample of four visualization experts (representing postdoctoral researchers and graduate students in visualization) participated in labeling. All labelers worked independently. 

\bstart{Labeling task}
Each labeler was asked to imagine that they were a visualization designer for a responsive visualization project, tasked with ranking a set of small screen design alternatives created by transforming the source. 
Their goal was to consider what would be an appropriate small screen design that would also preserve insights or takeaways conveyed in the desktop version as much as possible.

The study interface is shown in \autoref{fig:study-design}b. Each labeler completed 36 trials (6 desktop visualizations $\times$ 6 sets of 5 smartphone design candidates).
In each trial, the desktop visualization and five smartphone design candidates were shown, and labeler ranked the candidates by dragging and dropping them into an order. 
Trial order was randomized.

\bstart{Aggregating labels}
From the task, we collected human-judged rankings of 1,080 pairs each of which was ranked by three labelers.
To produce our training data set, we aggregated the three labels obtained from the three labelers of each pair into a single label representing the majority opinion, such that that for the $i$-th pair $\vx_i=(\vx_i^{(1)}, \vx_i^{(2)})$, the label $y_i$ is $1$ if $\vx_i^{(1)}$ is more likely to appear higher than $\vx_i^{(2)}$, and $-1$ otherwise.
\begin{equation}
    y_i= 
    \begin{cases}
        1,& \text{if } \vx_i^{(1)}\text{ more often appears higher than }\vx_i^{(2)}\\
        -1,& \text{ otherwise}
    \end{cases}
    \label{eq:label_aggregation}
\end{equation}
To avoid a biased distribution of training data as well as minimize the ordering effect within each pair, we randomized the order of pairs so that half of the pairs are labeled as $1$ and the other half as $-1$, which naturally sets the baseline training accuracy of 50\%.
\subsection{Model Description}
Prior approaches to visualization ranking problems (\eg~Draco-Learn~\cite{draco}, DeepEye~\cite{luo2018}) utilize ML methods that convert the ranking problem to a pairwise ordering problem, such as RankSVM (Support Vector Machine)~\cite{Herbrich1999} and the learning-to-rank model~\cite{burges2005}; we adopt a similar approach. 
A model, $f$, takes as input a pair of objects, $\vx = (\vx^{(1)}$, $\vx^{(2)})$, and returns their orders (\ie~either $\vx^{(1)}$ or $\vx^{(2)}$ ranks higher). 
\begin{equation}
    f(g(\vx^{(1)},\vx^{(2)})) = 
    \begin{cases}
        1,& \text{if } \vx^{(1)}\text{ appears higher in the ranking}\\
        -1,& \text{if } \vx^{(2)}\text{ appears higher in the ranking}
    \end{cases},
    \label{eq:rank_svm}
\end{equation}
where $g(\cdot,\cdot)$ is a mapping function that combines the features from a pair of objects.
We consider vector difference and concatenation for $g$.
Our models take two target views representing transformations of the same source view and return the one with higher predicted ranking, as depicted in \autoref{fig:prototype}.3.

\bstart{Features} We define the feature matrix $\mX \in \RR^{n \times d}$ where each row corresponds to a pair of target visualizations and columns represent the features (converted by $g$).
We use our proposed loss measures as the features (\autoref{tab:features}).
Aggregated features ($\mA$) refer to our three loss measures: identification, comparison, and trend loss, as described in Equations~\ref{eq:loss_entropy}, \ref{eq:loss_emd}, and \ref{eq:loss_trend} (\autoref{sec:loss_measures}). 
Disaggregated features ($\mD$) refer to the components of the aggregated features (\eg~the EMD value in each encoding channels for comparison loss).
We standardized all features.

\begin{table}[t]
    \centering
    \includegraphics[width=\columnwidth]{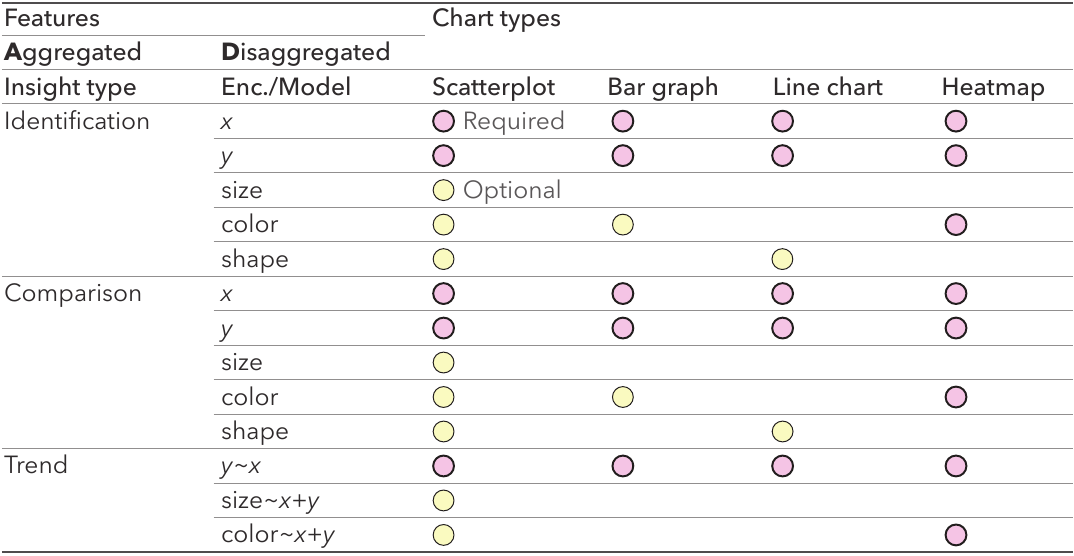}
    \caption{The set of features for our ML models by each chart type. These features are either concatenated or differentiated for each pair of targets.
    {\mA}ggregated features are the sum of the corresponding {\mD}isaggregated features. Pink, bold-bordered circles represent required features, and yellow, light-bordered circles optional encoding-specific features.}
    \label{tab:features}
\end{table}

\bstart{Model training} We train SVM with a linear kernel, K-nearest neighborhood (KNN) with $k=1,10$, logistic regression, decision tree (DT), and a Multilayer Perceptron (MLP) with four layers and 128 perceptrons per layer, similar to other recent applications of ML in data visualization (\eg~Hu \ea~\cite{hu2019}, Luo \ea~\cite{luo2018}). 
We also train ensemble models of DTs: random forest (RF) with 50, and 100 estimators, Adaptive Boosting (AB), and gradient boosting (GB).
Given the moderate number of observations (1,067) in our data set, we use leave-one-out (LOO) as a cross validation iterator to obtain robust training results.
We used Scikit-Learn~\cite{scikit} for training.

\bstart{Baselines} In addition to the natural baseline of 50\% (random), we include
two simple heuristic-based baselines to evaluate the performance of our models. 
The first baseline ($\mB$1) includes the changes in chart width and height between a target and its source, capturing an intuition about maintaining size and aspect ratio.
The second baseline ($\mB$2) is whether \textit{x} and \textit{y} axes are transposed, capturing an intuition that, of the strategies in our search space, transposing is the most drastic change.

\begin{figure}[t]
    \centering
    \includegraphics[width=\columnwidth]{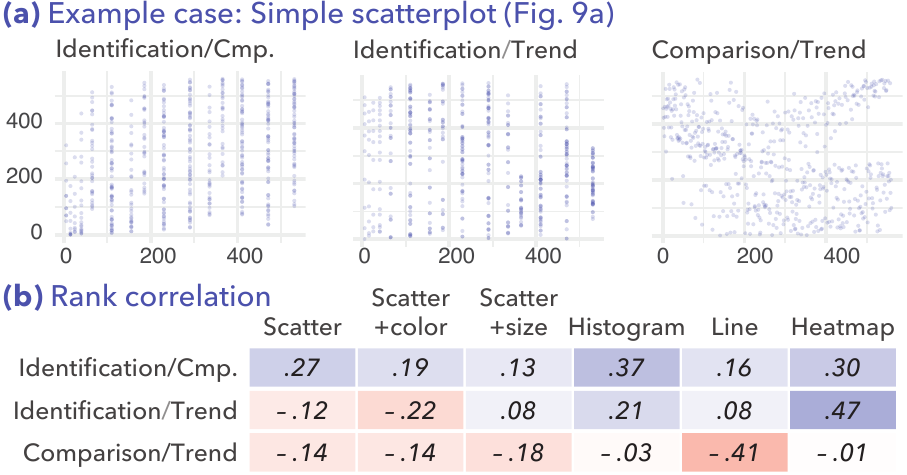}
    \caption{(a) Joint distributions of rankings of target views in each pair of aggregated features (the source visualization is shown in \autoref{fig:examples}a). (b) Kendall rank correlation coefficients for targets of our source views in \autoref{fig:examples}. Cmp (comparison).
    }
    \label{fig:correlation}
\end{figure}

\begin{table*}[h]
    \centering
    \includegraphics[width=\textwidth]{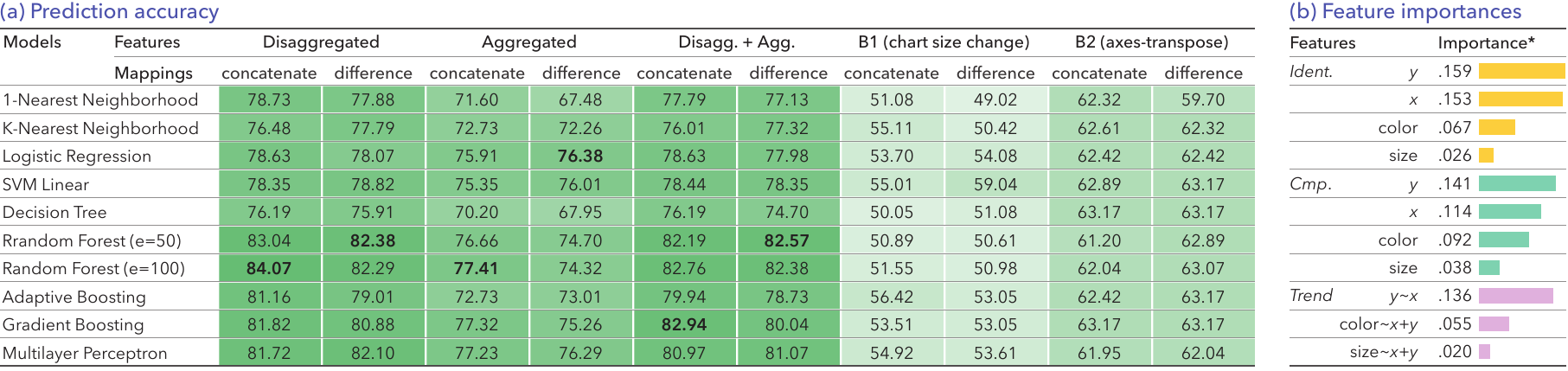}
    \caption{(a) Prediction accuracy of our models, averaged over LOO cross validation. 
    Other performance measures (AUC score and F1-score) appeared similarly to accuracy. 
    (b) Average importance of $\mD$isaggregated features ($g$ = difference) measured by impurity-based importance from training a random forest model (e = 50) 10 times.
    }
    \label{tab:result}
\end{table*}

\subsection{Results}\label{sec:results}
All the experimental materials, and files used for analysis are included in the supplementary materials, available at \url{https://osf.io/jcvbx}.

\subsubsection{Rank correlation between loss measures}
To ensure that our loss measures %(comparison, identification and trend) 
capture different information about transformations, we compute and inspect rank correlations between each pair of aggregated, and each pair of disaggregated measures. 
If two different loss measures produce highly similar rankings of target views, then one of them might be redundant.
Our measures tend to be orthogonal to each other (see \autoref{fig:correlation}), with Kendall rank correlation coefficients~\cite{kendall1948rank} between $-0.41$ and $0.47$.
The same pattern is observed for the disaggregated measures with overall correlation coefficients mostly between $-0.5$ and $0.5$ (see supplementary material).

When the chart type of a source view allows a few, limited responsive transformation strategies due to its own design constraints (\eg~line chart, heatmap), the correlation between measures appear slightly higher than the other chart types.
For example, it is often impossible to add a new encoding channel through aggregation or binning in a line chart.
This makes the line chart more sensitive to chart size changes, resulting in relatively higher negative rank correlation between comparison and trend loss.
Similarly, different binning levels in a heatmap can affect both one's ability to identify data points in different encoding channels and to recognize a trend implied by \textit{x} and \textit{y} on color channels (\ie~$e_\textit{color}~e_x+e_y$), leading to a slightly higher positive rank correlation between identification and trend loss.

\subsubsection{Monotonicity}\label{sec:monotonicity}
Ranking problems through pairwise comparison assume that the partial rankings used as input are consistent with the full ranking (monotonicity of rankings)~\cite{jamieson2011active,Herbrich1999}. 
In other words, we need to ensure that the partial pairwise rankings that we calculate based on aggregated expert labels can yield a monotonic full ranking. 
Comparison sorting algorithms can be used to determine whether a monotonic full ranking can be obtained from pairwise rankings, as a comparison sort will only result in a monotonic ranking if the principle of transitivity ($a>b \land b>c \implies a>c$) and connexity ($\forall a \text{ and } b, a \leq b \lor b \leq a$) hold.

To confirm whether our expert labels satisfy the monotonicity assumption, we first sort the five target views in each of our 108 trials, using the ten aggregated pairwise rankings as a comparison function. 
Next, we check whether each consecutive pair in the reproduced ordering conflicts with the aggregated expert labels, because if a pair in the reproduced ordering is not aligned with the aggregated label, that trial violates the monotonicity assumption. 
102 out of 108 trials ($94.44\%$) in our data set had fully monotonic orderings. 
Of the six orderings which are not fully monotonic, five are partially monotonic with only one misaligned pair each (out of the ten ordered pairs). 
The other non-monotonic ordering (a trial with line chart as the source view) had multiple conflicts; we dropped this ordering from our training data, resulting in 1,070 training pairs (1,067 unique training pairs).

\subsubsection{Training results}\label{sec:results_results}
\bstart{Model performance} 
Overall, our models with disaggregated ($\mD$) and aggregated features ($\mA$) achieved prediction accuracy greater than 75\% (\autoref{tab:result}a), showing the utility of our measures in ranking responsive design transformations.
Ensemble models (RF, AB, and GB) with $\mD$ features resulted in the highest overall accuracy (above 81\%) because they iterate over multiple different models and we have a relatively small number of features.
In particular, RF with $\mD$ and 100 estimators showed highest accuracy of 84\%.
Our neural network model (MLP) also provided comparable performance to the ensemble models.

Models with $\mD$ features in general obtained higher accuracy than $\mA$ features, and combining them ({$\mD+\mA$}) did not provide significant gain in accuracy. 
Although they had only three features, our models with $\mA$ features showed reasonable accuracy of up to $77.4\%$ ($g$ = concatenate) and $76.4\%$ ($g$ = difference). 
For mapping functions, concatenation performed slightly better than difference for our best performing models (RF). 

Our models all outperformed those with both baselines features ($\mB1$ and $\mB2$), indicating that our loss measures capture information that simple heuristics, such as changes in chart size or axes transposition, are unable to capture.
When we trained the best performing model (RF) with the features of only a single loss criterion (\eg~only trend), accuracy ranged from 52.6\% to 79.7\%, implying that our measures are more useful when combined than when used individually. 
As hy\-po\-thetical upper bounds for accuracy, training and testing the model on the same data set resulted in accuracy from 84\% (KNN) to 100\% (RF).

\bstart{Feature importance} To understand how the different loss measures function in our models, we inspected the importance of each disaggregated feature (mapping function $g$ = difference) using the impurity-based importance measure (average information gain) by training a random forest model with 50 estimators (average over 10 training iterations).
As shown in \autoref{tab:result}b, features related to position encodings ($x$, $y$, $e_x \sim e_y$) in general seem to have higher importance, which makes sense given their ubiquity in our sample.
%This makes sense in light of the well known primacy of position encodings in visualization judgments~\cite{kim2018,mackinlay1986automating}.
%The lower importance of size and color encodings and trend models with those encodings is not surprising given that they appear sparsely in our data set.

\bstart{Predicted rankings of example cases}
Using the best prediction model (RF with 100 estimators), we predicted the rankings of example cases described in \autoref{sec:examples}.
Transformations from the simple scatterplot example (\autoref{fig:examples}a) are ranked as: Ta1 (resizing), Ta2 (transposing axes), Ta4, Ta3 (binning, resizing, aggregation), and Ta5 (binning, resizing, aggregation, mark type change). 
Ta1 appears higher in ranking than Ta2 because Ta2 has higher trend loss, while Ta1 slightly sacrifices comparison loss.
Ta4 is ranked in a higher position than Ta5 because Ta5 has higher comparison and trend loss.
Responsive transformations from the histogram example (\autoref{fig:examples}b) are ranked as: Tb2 (transposing axes, resizing), Tb3 (transposing axes, resizing), Tb1 (resizing), Tb5 (resizing, changing bin size), and Tb4 (resizing, changing bin size). 
The transposed views (Tb2 and Tb3) are ranked higher than Tb1 probably because the model has more emphasis on comparison loss as the feature importance (\autoref{tab:result}b) shows.
The ordering between Tb5 and Tb4 can be backed by the smaller trend loss of Tb5 while the difference in comparison loss between them appears subtle.

\section{Discussion and Future Work}
\subsection{Extending and Validating Our Preservation Measures}
We devised a small set of measures for three common low-level tasks in visualization use and found that they can be used to build reasonably well-performing ML models for ranking small screen design alternatives given a large screen source view.
Our measures are not strongly correlated, and removing some of the measures results in lower predictive accuracy.
However, there are other forms of prominent task-oriented insights that could extend our approach if approximated well, such as clustering data points or identifying outliers.
As our measures lose information by processing rendered values, future work could estimate task-oriented insights with different methods, such as extracting and directly comparing image features from rendered visualizations.

There are also opportunities to strengthen and extend our measures through human subject studies.
These include more formative research with mixed methods to understand heuristics and other strategies that visualization authors and users employ to reason about how well a design transformation preserves important takeaways.
In addition, future work could conduct perceptual experiments that more precisely estimate human baselines for identification, comparison, and trend losses.  
We also used simple approximations of perceptual differences in position, size, and color channels which could be improved through new experiments specifically designed to understand how perception is affected on smaller screen sizes, adding to work like examining task performance on smaller screens by different chart types~\cite{brehmer2020} and comparing task performance between small and large screens~\cite{Blascheck2019,Blascheck2021}.
A limitation is that our experiment was conducted on desktop devices. Future work could test on mobile devices, as well as explore mobile-first design contexts, as our measures are designed to be symmetric.

\subsection{Responsive Visualization Authoring Tools} 
Our work demonstrates how task-oriented insight preservation can be used to rank design alternatives in responsive visualization. 
To do so, we formulated and evaluated our insight preservation measures on a search space representing common responsive visualization design strategies and mark-encoding combinations. 
However, our work should be extended in several important ways to support a responsive visualization authoring use case. 

First, while more drastic encoding changes than those supported by our generator are rare in practice~\cite{kim2021}, this might be because responsive visualization authoring is currently a tedious process and authors satisfice by exploring smaller regions of the design space. 
There are many strategies that could be added to a search space like the one we defined, and used to evaluate our measures as well as to learn more about how authors react when confronted with more diverse sets of design alternatives.
For example, while we mainly consider single-view, static visualizations, many communicative visualizations employ multiple views and interactivity~\cite{Segel2010,Hullman2011,hullman2013deeper}. 
Ideally a responsive visualization recommender should be able to formulate related strategies (\eg~rearranging the layout of multiple views, omitting an interaction feature, editing non-data ink like legends).
Recommenders may need to consider further conditions such as consistency constraints for multiple views~\cite{qu2016}, effectiveness of visualization sequence~\cite{kim2017graphscape}, semantics of composite visualizations~\cite{javed2012}, and effectiveness of interactive graphical encoding~\cite{saket2018}.
As indicated in Kim~\ea~\cite{kim2021}, loss measures should be able to address concurrency of information because rearranging multiples views (\eg~serializing) can make it difficult to recognize data points at the same time on small screen devices.
In addition, they should also account for loss of information that can only be obtained via user interaction (\eg~trend implied by filtered marks).

We envision our measures, and similar measures motivated to capture other task-oriented insights, being surfaced for an author to specify preferences on in a semi-automated responsive visualization design context. 
Because what our measures capture is relatively interpretable, authors may find it useful to customize them for certain design tasks, such as prioritizing one measure or changing how information is combined to capture identification, comparison, or trend loss.
This is a strength of our approach relative to using a more ``black-box'' approach where model predictions might be difficult to explain. 

\subsubsection{Extending ML-based approaches} 
The human labelers in our experiment, including the authors, seemed to at times use strategies or heuristics such as preferring non-transposed views in their rankings or trying to minimize changes to aspect ratio for some chart types.
However, models with our loss measures as features perform better than heuristic approaches like detecting axes-transpose and chart size changes, implying that task-oriented insights may be the right level at which to model rankings.
As an extension, future work might learn pre-defined costs for different transformation strategies to reduce the time complexity of evaluating task-oriented insights preservation, similar to the approach adopted by Draco-Learn~\cite{draco} which obtained costs for constraint violation.
Learning such pre-defined costs may also enable better understanding how each responsive design strategy contributes to changes in task-oriented insights.
An alternative approach could be to use our loss measures as cost functions and optimize different strategies to reduce them as MobileVisFixer~\cite{wu2020mobilevisfixer} fixes a non-mobile-friendly visualizations for a mobile screen by minimizing heuristic-based costs.
As recent deep learning models~\cite{wu2021,ma2020} have performed well in visualization ranking problems, future work may further elaborate on those models.
In doing so, one could combine our measures with image features (\eg~ScatterNet~\cite{ma2020}) or chart parameters (\eg~aspect ratio, orientation~\cite{wu2021}).

As noted in \autoref{sec:monotonicity}, there were a few partial and not fully monotonic orderings in our data set.
A better model might ignore this assumption and try to identify highly recommendable transformations or classify them into multiple ordinal classes, yet this might come up with lower interpretability about recommendations due to a lack of explicit ordinal relationship between transformations.

\subsection{Generalizing Our Measures to Other Design Domains}
Our approach to task-oriented insight preservation is likely to be useful in visualization design domains beyond responsive visualization, like style transfer and visualization simplification, although other domains may also require different transformation strategies.
Style transfer, for instance, may involve techniques like aggregation but is more likely to change visual attributes of marks or references. 
While our loss measures are designed to be low-level enough to apply relatively generically to visualizations, their precise formulation and the combination strategy might warrant changes in other domains.  
For example, in visualization simplification, minimizing trend loss is likely to be more important than preserving identification and comparison of individual data points. 
Style transfer often focuses on altering color schemes, size scales, or mark types~\cite{Harper2014} which can result in different discriminability distributions, so it might put more emphasis on comparison loss. 
\section{Conclusion}
Responsive visualization transformations often alter task-oriented insights obtainable from a transformed view relative to a source view.  
To enable automated recommenders for iterative responsive visualization design, we suggest loss measures for identification, comparison, and trend insights.
We developed a prototype responsive visualization recommender that enumerates transformations and evaluates them using our measures.
To evaluate the utility of our measures, we trained ML models on human-produced orderings that we collected, achieving accuracy of up to $84.1\%$ with a random forest model.

% %% if specified like this the section will be committed in review mode
\acknowledgments{Jessica Hullman thanks NSF (\#1907941) and Adobe.}

\bibliographystyle{abbrv-doi-hyperref-narrow}

\bibliography{main}
\end{document}